\documentclass[10pt,a4paper,twocolumn,superscriptaddress]{revtex4}

\usepackage[a4paper,
            top=2.54cm,
            bottom=2.54cm,
            left=1.5cm,
            right=1.5cm]{geometry}

\usepackage{textcomp}
\usepackage{color}
\usepackage{amsmath}
\usepackage{mathrsfs}
\usepackage{graphicx}
\usepackage{subfig}
\usepackage{epsfig}
\usepackage{physics}
\usepackage{amssymb}
\usepackage{hyperref}
\usepackage{todonotes}
\usepackage[final]{pdfpages}
\usepackage{float}

\newcommand{\micron}{\textmu{m}}

\usepackage[squaren,Gray]{SIunits}
\usepackage{pgfplots}
\pgfplotsset{compat=newest}
\usepgfplotslibrary{groupplots}
\usepgfplotslibrary{polar}
\usepgfplotslibrary{smithchart}
\usepgfplotslibrary{statistics}
\usepgfplotslibrary{dateplot}
\usepgfplotslibrary{ternary}

\begin{document}

\title{Electrically Pumped Quadratic-Soliton Mode-Locked Laser\\ in Lithium Niobate Nanophotonics}

\author{Benjamin K. Gutierrez}
\altaffiliation{These authors contributed equally.}
\affiliation{Department of Applied Physics and Materials Science, California Institute of Technology, Pasadena, California 91125, USA}

\author{Nicolas Englebert}
\altaffiliation{These authors contributed equally.}
\affiliation{Department of Electrical Engineering, California Institute of Technology, Pasadena, California 91125, USA}

\author{Yingchu Xu}
\altaffiliation{These authors contributed equally.}
\affiliation{Department of Electrical Engineering, California Institute of Technology, Pasadena, California 91125, USA}

\author{Selina~Zhou}
\altaffiliation{These authors contributed equally.}
\affiliation{Department of Electrical Engineering, California Institute of Technology, Pasadena, California 91125, USA}

\author{Maximilian Shen}
\affiliation{Department of Electrical Engineering, California Institute of Technology, Pasadena, California 91125, USA}

\author{Mahmood Bagheri}
\affiliation{Jet Propulsion Laboratory, California Institute of Technology, Pasadena, California 91109, USA}

\author{Alireza Marandi}
\email{marandi@caltech.edu}
\affiliation{Department of Applied Physics and Materials Science, California Institute of Technology, Pasadena, California 91125, USA}
\affiliation{Department of Electrical Engineering, California Institute of Technology, Pasadena, California 91125, USA}

\begin{abstract} %
\leftskip=2.5cm \rightskip=2.5cm
\noindent
Femtosecond mode-locked lasers are the workhorse of modern photonics, enabling optical atomic clocks, frequency combs, precision spectroscopy, and ultrafast signal processing\,\cite{keller_ultrafast_2021}.
Bringing them onto nanophotonic chips promises miniaturized ultrafast optical systems, yet combining femtosecond pulse generation with electrical pumping remains a major challenge, rooted in a trade-off between the speed and energy efficiency of conventional mode-locking mechanisms.
Kerr-based mechanisms offer nearly instantaneous response but typically require high intracavity peak powers\,\cite{qiu_high-pulse-energy_2026, singh_towards_2020} that are difficult to achieve with nanophotonic-compatible semiconductor gain.
Conversely, semiconductor saturable absorbers enable strong nonlinear responses at low optical energies but are limited by their slow carrier dynamics\,\cite{karin_ultrafast_1994, green_fast_2011}.
Here we break this trade-off by introducing quadratic-soliton mode-locking based on a quadratic nonlinear interferometer (QNLI): an effective saturable absorber built from cascaded second-order nonlinearity in a periodically poled lithium niobate nanophotonic circuit, with a measured sub-20\,fs response time.
Through hybrid integration with an electrically pumped semiconductor gain element, we demonstrate a self-starting quadratic-soliton-based mode-locked laser with femtojoule pulse energies, operating in single- and multiple-soliton states.
Quadratic solitons arise through the QNLI’s dissipative and dispersive nonlinear response, which we theoretically confirm using a modified complex Ginzburg--Landau equation in agreement with measurements.
By combining soliton formation with intracavity electro-optic modulation, we generate 750\,fs pulses with flat-top spectra.
Our engineerable, ultra-energy-efficient, electrically powered mode-locked laser provides the critical source for fully integrated ultrafast photonic circuits in lithium niobate nanophotonics that natively offers a wide range of functionalities\,\cite{zhu_integrated_2021} including electro-optic control\,\cite{wang_integrated_2018}, quadratic wavelength conversion\,\cite{wang_ultrahigh-efficiency_2018}, energy-efficient pulse compression\,\cite{gray_two-optical-cycle_2026}, and spectral broadening\,\cite{sekine_multi-octave_2025}.
\end{abstract}

\maketitle
Femtosecond mode-locked lasers (MLLs) have enabled numerous modern technologies and extended the boundaries of experimental research\,\cite{keller_ultrafast_2021}. By generating trains of ultrashort optical pulses in the time domain, which typically correspond to controllable frequency combs in the spectral domain\,\cite{diddams_optical_2020}, they have fortified the foundation of nonlinear optics, attosecond physics\,\cite{krausz_attosecond_2009}, as well as high-precision science such as time metrology\,\cite{ludlow_optical_2015} and spectroscopy\,\cite{picque_frequency_2019}. These advances have motivated substantial efforts to reduce the size, cost, and complexity of MLLs\,\cite{chang_integrated_2022} and that of the photonic systems driven by them\,\cite{siew_review_2021,zhu_integrated_2021}.

Among the nanophotonic platforms investigated to this end, thin-film lithium niobate (TFLN) stands out for combining strong second-order ($\chi^{(2)}$) nonlinearity and electro-optic response\,\cite{hu_integrated_2025} with dispersion-engineering capabilities\,\cite{zhu_integrated_2021}, enabling modulators\,\cite{wang_integrated_2018}, time lenses\,\cite{yu_integrated_2022}, optical parametric amplifiers\,\cite{ledezma_intense_2022} and oscillators\,\cite{Lu26_OPO_Review}, quantum-state generation\,\cite{nehra_few-cycle_2022,labbe_thin-film_2025}, and spectral broadening\,\cite{sekine_multi-octave_2025}, among others. 
Together, these capabilities lay the foundation for monolithic ultrafast photonic integrated circuits (PICs), creating a growing demand for integrated femtosecond MLLs in TFLN.

The most common femtosecond sources in nanophotonic platforms, including Kerr combs\,\cite{herr_temporal_2014, kippenberg_dissipative_2018, pasquazi_micro-combs_2018} and electro-optic combs\,\cite{hu_high-efficiency_2022}, have required optical driving or pumping. 
Electrically pumped semiconductor lasers can serve as the optical pump for these femtosecond comb-generation schemes\,\cite{yu_integrated_2022, Shen2020, hu_chip-scale_2025}; however, semiconductor laser operation remains independent of the comb-generation mechanism.  
More recently, Mamyshev oscillator-based MLLs have also been demonstrated using optically pumped laser gain\,\cite{qiu_high-pulse-energy_2026}. 
In these systems, the external driving laser typically imposes constraints on chip-scale integration and scalability, and the mechanism for femtosecond pulse generation imposes constraints on the wall-plug efficiency, while self-starting and turn-key operation remain challenging.

These constraints make MLLs based on well-developed, electrically pumped semiconductor gain elements highly desirable as scalable and efficient femtosecond sources in nanophotonics. 
However, most nanophotonic semiconductor MLLs have so far demonstrated picosecond pulse durations\,\cite{guo_ultrafast_2023,hermans_high-pulse-energy_2021,wang_passive_2026}, despite broadband operation\,\cite{Zhang_Broadband_III-V_26}. For passively mode-locked lasers, this has been associated with the relatively slow recovery of semiconductor saturable absorbers\,\cite{karin_ultrafast_1994, green_fast_2011}. 
On the other hand, effective saturable absorbers based on Kerr nonlinearities can enable ultrafast operation, but they typically require substantial peak powers\,\cite{singh_towards_2020, qiu_high-pulse-energy_2026} beyond the reach of semiconductor gain elements suited to nanophotonic integration.
Moreover, in many semiconductor laser systems, strong competing nonlinearities in the gain medium counteract the desired passive mode-locking mechanism at high peak powers\,\cite{resan_dispersion-managed_2004,delfyett_high-power_1992}.
 
Lithium niobate nanophotonics offers fertile ground for exploring electrically pumped semiconductor MLLs promising to address these challenges.
In contrast to table-top demonstrations using bulk crystals, where laser mode locking with quadratic nonlinearity has remained sparse\,\cite{stankov_mirror_1988}, the strong quadratic nonlinearity of LN combined with dispersion engineering and quasi-phase matching engineering in nanophotonics enables laser mode locking regimes that have not been accessible in other photonic platforms. In particular, the solitonic solutions associated with quadratic resonators\,\cite{englebert_temporal_2026} are of substantial interest, which, when combined with electrically pumped gain elements, offer the potential for energy-efficient self-starting and stable MLL operation.

The engineerable effective Kerr nonlinearity arising from the cascaded quadratic nonlinearity in LN nanophotonics provides unique features for integrated laser mode locking. 
First, the sign of the effective Kerr can be chosen based on the sign of the phase mismatch ($\Delta k$) using periodic poling, enabling soliton formation in cavities with either normal or anomalous dispersion. 
Second, this nonlinearity can be introduced to a specific location through periodic poling, while the intrinsic Kerr is present everywhere. 
Finally, the magnitude of the effective Kerr nonlinearity can be tailored and can be much stronger than that of the intrinsic Kerr effect, owing to the strong quadratic nonlinearity in LN (SI section II, Fig.~S3).
Because of this intrinsically strong second-order nonlinearity, the QNLI intensity-dependent complex reflectivity can reach beyond that of intrinsic-Kerr-based implementations (SI section III, Fig.~S4).
Hence, the effective Kerr can be designed to work in relatively low peak powers and be the dominant nonlinearity of the system, which is particularly important in the presence of undesired competing nonlinearities in the semiconductor gain element\,\cite{resan_dispersion-managed_2004,delfyett_high-power_1992}.

Laser mode-locking through quadratic nonlinearity in LN nanophotonics also opens a regime that combines ultra-energy-efficient operation with ultrafast nonlinear response. This combination has been inaccessible in previous demonstrations of various mode-locking schemes (SI section V, Fig. S6, Table 1). Achieving an ultrafast nonlinear response alongside ultra-energy-efficient operation lowers the mode-locking threshold and enables higher repetition rate harmonic mode-locking at modest average powers. Such properties could enable milliwatt-level power per comb line at repetition rates in the hundreds of gigahertz. Alternatively, the quadratic nonlinearity can be engineered to mode-lock at higher pulse energies, enabling fundamental mode-locking at high peak powers. This flexibility further opens up the design space for nanophotonic MLLs across the board.

In this work, we show, for the first time, a turn-key semiconductor laser in LN nanophotonics that is mode-locked in a quadratic-soliton regime based on the effective Kerr nonlinearity. 
The soliton-MLL is formed by integrating a quadratic nonlinear interferometer (QNLI) in an LN nanophotonic circuit with an electrically pumped reflective semiconductor optical amplifier (RSOA). Through dispersion engineering and quasi-phase matching in LN\,\cite{jankowski_ultrabroadband_2020,ledezma_intense_2022}, the QNLI acts as an artificial saturable absorber (SA) \cite{keller_ultrafast_2021}, providing intensity-dependent complex reflection leading to the formation of dissipative solitons, verified by a modified complex Ginzburg-Landau equation (CGLE). 
\begin{figure*}
    \centering
    \includegraphics[width=1\linewidth]{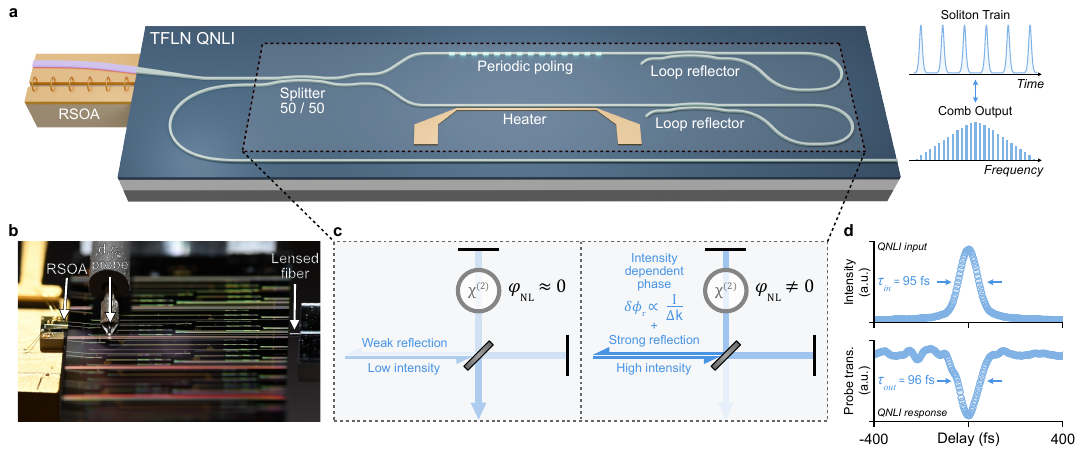}
    \caption{\textbf{Quadratic-Soliton Mode-Locked Laser Concept.} 
    \textbf{a}, Schematic of the nanophotonic QNLI implementation and its hybrid integration with a reflective semiconductor optical amplifier (RSOA) forming an electrically pumped passively mode-locked laser (MLL). 
    \textbf{b}, Image of the QNLI-based electrically pumped passively mode-locked laser.
    \textbf{c}, The operating principle of the quadratic nonlinear interferometer (QNLI) providing a peak-power-dependent reflection in the laser cavity.
    \textbf{d}, Measured ultrafast response of the QNLI. Input-pulse autocorrelation (top) and pump-probe transmission dip (bottom), with 95\,fs and 96\,fs FWHMs, indicate a $\sim$14\,fs response time (SI section IV, Fig. S5).}
    \label{Fig1:Concept}
\end{figure*}

\section*{Results}
\noindent
Figure\,\ref{Fig1:Concept}a shows the integrated semiconductor mode-locked laser architecture, which is based on a nanophotonic QNLI in the form of a nonlinear Michelson interferometer\,\cite{ouellette_pulse_1986,ouellette_passive_1988,singh_towards_2020,singh_silicon_2024} operating in reflection inside the laser cavity: the higher the input intensity, the stronger the reflection. Figure\,\ref{Fig1:Concept}b shows an image of the device. The nanophotonic circuit of the QNLI consists of a balanced coupler (50/50) splitting the power into two arms, each of which is terminated by loop reflectors. 
The bottom arm includes a heater for controlling the operating point of the QNLI.
The top arm includes a periodically poled section that introduces an engineerable \textit{effective Kerr response} that yields an intensity-dependent phase, $\varphi_\text{nl} \propto [\chi^{(2)}]^2/\Delta k$\,\cite{desalvo_self-focusing_1992,stegeman_2_1996}. 
The magnitude of reflection from the QNLI inside the laser cavity depends on this nonlinear phase, which also imposes an intensity-dependent phase on the reflected field, as illustrated in Fig.\,\ref{Fig1:Concept}c. Thus, unlike typical SAs, the QNLI additionally provides a strong intensity-dependent phase. The interferometer's static phase can be set at the point of no reflection, where low input intensities result in $\varphi_\text{nl}\approx0$ and $R\approx0$. At this setpoint, however, high input intensities result in strong reflection, reaching $\sim$100\% when $\varphi_\text{nl}=\pi/2$.
The near-instantaneous nonlinearity of the QNLI combined with its dispersion engineering provides an ultrafast response. We measured this response using a pump-probe technique (SI section IV), which indicates a resolution-limited response time of $\sim$14 fs, obtained by deconvolving the input-pulse autocorrelation from the pump-probe trace (Fig.\,\ref{Fig1:Concept}d).

\section*{Quadratic soliton mode-locking}
\noindent 
The complex intensity-dependent reflectivity of the QNLI can lead to dissipative soliton formation in the laser cavity, which we study using a modified CGLE (SI section VI.A) \cite{keller_ultrafast_2021,bache_nonlocal_2007,englebert_temporal_2026, nie_photonic_2022}. The field envelope $A$ oscillating at a frequency $\omega_0$ is modeled by
\begin{align}
t_R\frac{\partial A_m}{\partial T}
&= \left(\frac{gL}{2}-\frac{\alpha L}{2}+D\frac{\partial^2}{\partial \tau^2}\right) A_m, \label{eq:qnli}\\[-1.2em]
A_\text{m+1}&= (-\theta +(1-\theta)e^{i\varphi}) A_\text{m}+ \overbrace{\kappa^2L_2^2\theta^2[A_\text{m}^2\otimes I ]A_\text{m}^*}^{\chi^{(2)}\text{ section}},\nonumber
\end{align}
where $t_R$ is the roundtrip time, $T$ is a slow time describing the roundtrip-to-roundtrip evolution with $m$, the roundtrip number, and $\tau$ is a reference time frame traveling at the group velocity at $\omega_0$. $^*$ denotes the complex conjugate, $g$ describes the saturated semiconductor gain, $\alpha$ is the total roundtrip loss in intensity, and $L$ is the total cavity length. $D = D_g-i\beta_2L/2$, where $D_g$ is the gain/cavity bandwidth, corresponding to the combination of the gain and cavity spectral response, and $\beta_2$ is the group velocity dispersion (GVD) parameter at $\omega_0$. 
The second equation is the nonlinear transfer function of the QNLI [SI section I.C, Eq.~(S9)], in which $\theta$ is the splitter's transmission coefficient in power, and $\varphi$ is a phase bias accounting for an extra linear phase difference between the two QNLI arms. The second term describes the quadratic nonlinear section: $\kappa\propto\chi^{(2)}$ is its effective second-order nonlinear coefficient, $L_2$ its length, and $\otimes$ denotes the convolution with the nonlocal kernel $I$\,\cite{bache_nonlocal_2007,valiulis_propagation_2011,leo_walk-off-induced_2016,englebert_temporal_2026}.
The kernel depends on the phase mismatch, the GVD at both $\omega_0$ and $2\omega_0$, as well as the group velocity mismatch (GVM) between these two frequencies. 
\begin{figure*}
    \centering
    \includegraphics[width=1\linewidth]{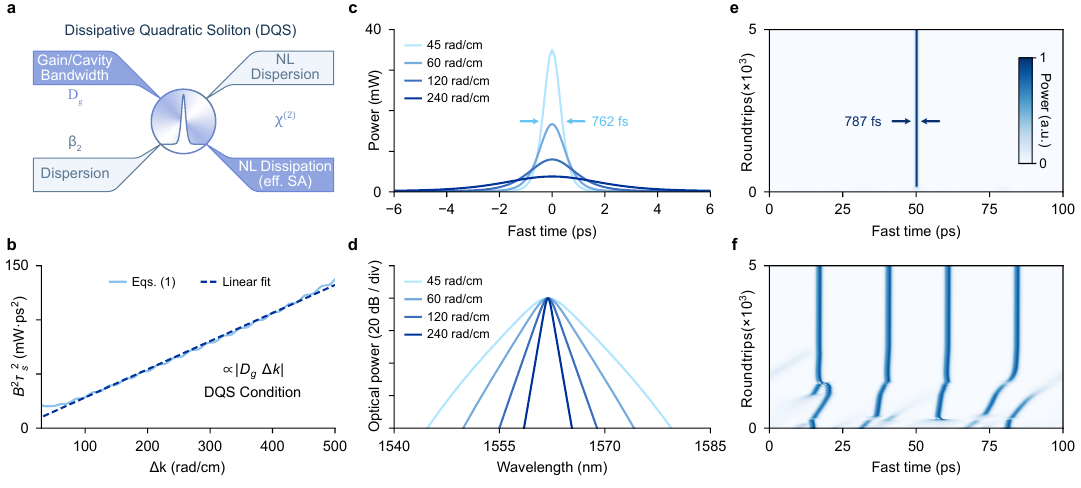}
    \caption{ \textbf{Theoretical analysis of quadratic soliton passive mode-locking.} 
    \textbf{a},\,Dissipative quadratic solitons (DQS) emerge due to both the real (dissipative) and imaginary (dispersive) response of the QNLI balanced with the gain bandwidth and chromatic dispersion, respectively. 
    \textbf{b},\,The numerical integration of the modified CGLE (Eqs.\,\eqref{eq:qnli}) in the bandwidth-limited regime confirms the soliton condition, $B^2\tau_s^2 \propto |D_g\Delta k|$, when the phase mismatch is varied. 
    \textbf{c}, Dissipative soliton temporal profile and \textbf{d}, spectrum for different phase mismatch. 
    \textbf{e},\,Simulation of the full lumped-element model (SI section VI.B), starting from quantum noise with $\varphi = -\pi/10$\,rad and $\Delta k$ = 45 rad/cm, reveals the spontaneous formation of a dissipative quadratic soliton, in agreement with the theoretical prediction (b-d). 
    \textbf{f}, The system evolves towards harmonic mode-locking when the phase bias is changed to $\varphi = -2.5$\,rad. 
    }
    \label{Fig2:Simulations}
\end{figure*}
For large phase mismatch ($\Delta k L_2 \gg 1$), this last term introduces an effective Kerr response proportional to $\kappa^2 L_2I/\Delta k$ through cascaded quadratic nonlinear effects [SI section II, Eq.~(S10)].
Yet, as $I$ is a complex function, the effective Kerr, and thus the QNLI's reflectivity, has both a real and an imaginary part. This plays two important roles. 
On the one hand, the real component is responsible for the QNLI's strong saturable absorption. 
The imaginary part, on the other hand, provides a phase response which can compensate for the system's dispersion. Since the sign of this phase response is given by the phase mismatch, the QNLI admits bright solitons in both dispersion regimes. 

Dissipative \textit{Kerr} soliton mode-locking has been thoroughly investigated over the last three decades\,\cite{kivshar_optical_2003,keller_ultrafast_2021,grelu_dissipative_2012}. 
Here, in sharp contrast, the formation of dissipative \textit{quadratic} solitons (DQSs)\,\cite{buryak_optical_2002,nie_photonic_2022,englebert_temporal_2026} involves a distinct double balance, in addition to the gain-dissipation equilibrium achieved under lasing conditions: the real and imaginary parts of the QNLI's response compensate for the system's gain bandwidth and chromatic dispersion, respectively. This is illustrated in Fig.\,\ref{Fig2:Simulations}a, where the nonlinear dispersion (phase response) and nonlinear dissipation (intensity response) of the QNLI balance the linear dispersion and gain bandwidth in the cavity, respectively.

In the steady state and for large phase mismatch, Eq.\,\eqref{eq:qnli} admits bright DQS solutions of the form $A_s = B\,\text{sech}^{1+ic_s}(\tau/\tau_s)$ where $B$ and $\tau_s$ are the soliton amplitude and duration, respectively, and $c_s$ is the chirp parameter (SI section VII.A).
As in conservative Kerr nonlinear media, the soliton amplitude and duration are linked to each other by the soliton condition\,\cite{kivshar_optical_2003}. Here, this condition reads $B^2\tau_s^2 \propto |D\Delta k|$ (SI section VII.B). 
For a sufficiently broad gain/cavity bandwidth, $|D|\to |\beta_2L/2|$ and, since $\gamma_\text{eff}\sim 1/\Delta k$, one recovers the well-known conservative bright Kerr soliton condition if the dispersion is anomalous\,\cite{kivshar_optical_2003}. However, for DQSs, this condition can remain valid in the normal dispersion regime ($\beta_2>0$), provided that the phase-mismatch is negative ($\Delta k <0$), which can be achieved through quasi-phase matching in lithium niobate (SI section VII.C, Fig.~S8).

For the case of $|D|\to D_g$, as in our experiments, the impact of the GVD is negligible, and the soliton duration is determined by the balance between the linear and nonlinear dissipative terms, i.e., the gain/cavity bandwidth, and the magnitude of the QNLI saturable absorption, which is proportional to the effective Kerr coefficient. In fact, due to the strong saturable absorption and the double balance provided by the QNLI (Fig.\,\ref{Fig2:Simulations}a), DQS can form irrespective of the sign of $\Delta k \cdot \beta_2$ (SI section VII.C, Fig.~S7a). This is corroborated by the numerical integration of Eqs.\,\eqref{eq:qnli}, revealing the existence of DQS solutions (Figs.\,\ref{Fig2:Simulations}b-d). 
To assess the validity of Eqs.\,\eqref{eq:qnli}, we further integrate the full lumped-element model describing our system (SI section VI.B), starting from quantum noise, at $\varphi = -\pi/10$\,rad and $\Delta k$ = 45 rad/cm. The 2D colormap, plotted in Fig.\,\ref{Fig2:Simulations}e, highlights self-starting fundamental soliton mode-locking operation whose duration is in excellent agreement with the prediction of Eqs.\,\eqref{eq:qnli} (Fig.\,\ref{Fig2:Simulations}c). When the phase bias is set to $\varphi = -2.5$\,rad, the system evolves towards harmonic mode-locking (Fig.\,\ref{Fig2:Simulations}f). 

\section*{Quadratic-Soliton Mode-Locked Laser Operation}
\begin{figure*}[tp]  
    \centering
    \hspace{-5mm}
    \includegraphics[width=183mm, keepaspectratio]{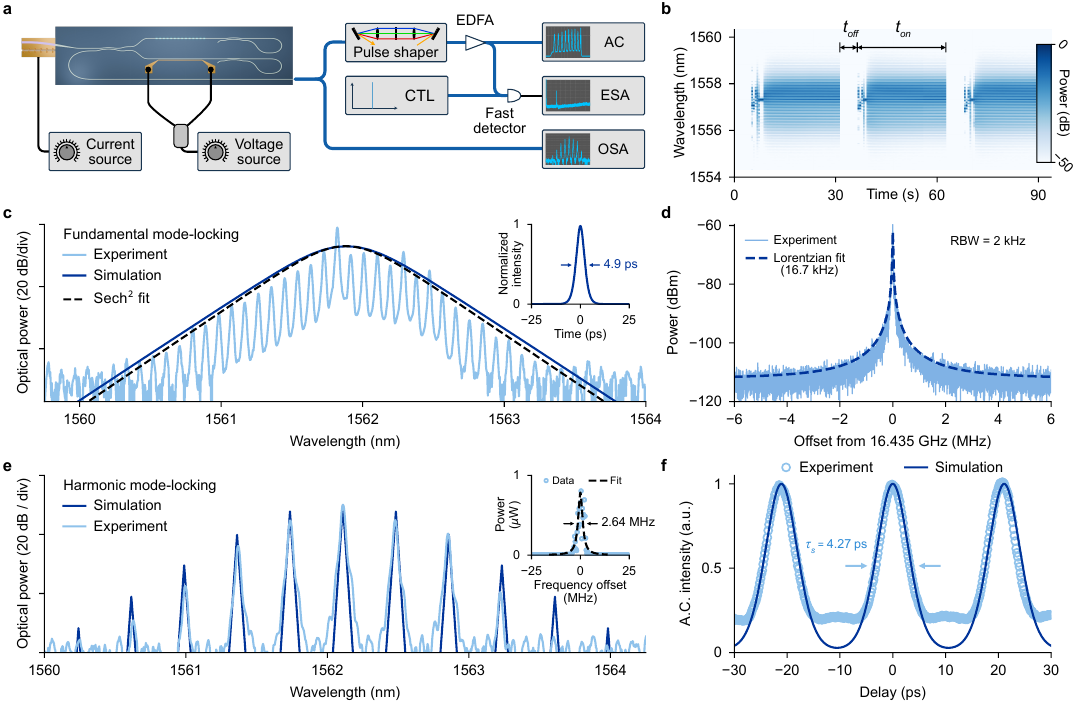}
    \caption{\textbf{Measurement results of the quadratic-soliton-mode-locked laser operation}
        \textbf{a}, Schematic of the experimental setup. EDFA, erbium-doped fiber amplifier; OSA, optical spectrum analyzer; A.C., Intensity Autocorrelator; ESA, electrical spectrum analyzer; CTL, continuously tunable laser.
        \textbf{b}, Spectra of the MLL as the current is switched on and off, demonstrating self-starting behavior.
        \textbf{c}, Measured spectrum of the fundamental passive mode-locking regime. A sech$^2$ fit and the simulated spectral envelope are also plotted. The inset shows the simulated pulse. 
        \textbf{d}, Measured RF linewidth of the passively mode-locked laser with a Lorentzian fit of 16.7 kHz FWHM, obtained with a different RSOA (Methods).
        \textbf{e}, Measured and simulated spectra of the harmonic passive mode-locking regime. Inset shows a heterodyne beat note measurement with a Lorentzian fit of 2.64\,MHz FWHM. 
        \textbf{f}, Measured and simulated intensity autocorrelation of the harmonic passive mode-locked state, with a 4.27 ps deconvolved pulse duration. 
    }    
    \label{Fig3:PMLL}
\end{figure*} 
\begin{figure*}[tp]  
    \centering
    \hspace{-5mm}
    \includegraphics[width=183mm, keepaspectratio]{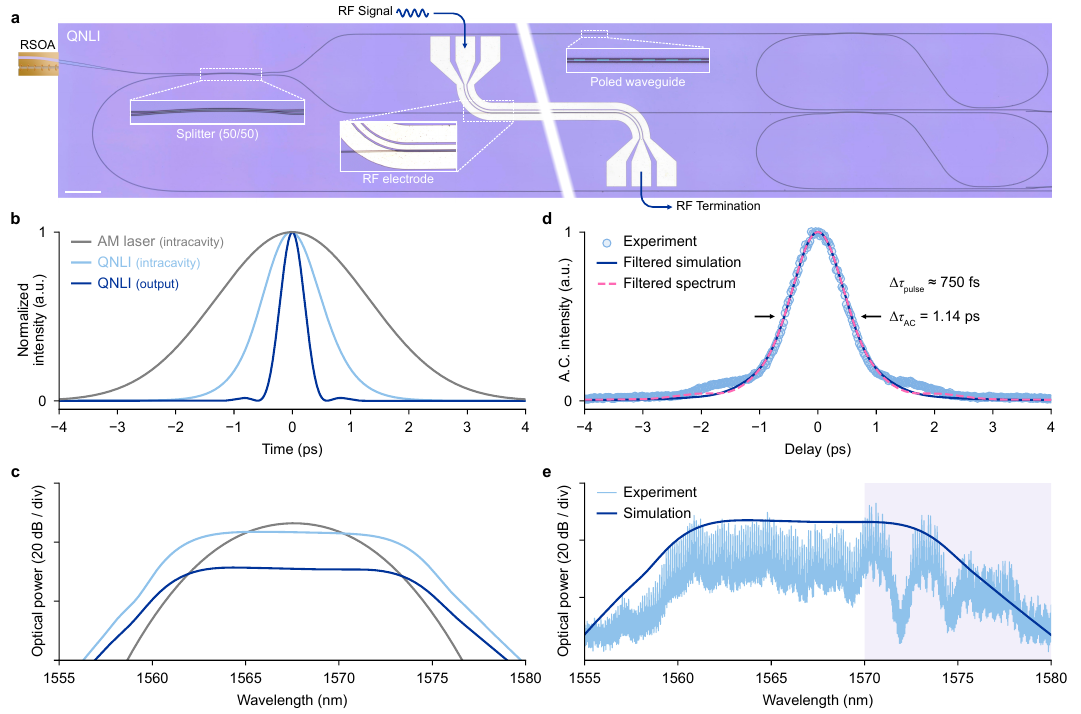}
    \caption{
    \textcolor{black}{
        \textbf{Femtosecond MLL operation} 
        \textbf{a}, Schematic of the MLL (not to scale) including the microscope image of the QNLI nanophotonic chip in which the EO modulator is driven for femtosecond operation. The heater shown in Fig.\,\ref{Fig1:Concept}b is replaced by fast radiofrequency (RF) electrodes. The input taper on the left-hand side goes to the RSOA (illustration not to scale). 
        \textbf{b}, Simulated temporal and \textbf{c}, simulated spectral profiles. Without a $\chi^{(2)}$ section, TFLN PIC corresponds to an amplitude-modulated MLL, producing Gaussian pulses (gray). In contrast, the QNLI poled section leads to soliton mode-locking (light blue). Dispersion compensation further shortens the output pulse duration (dark blue).
        \textbf{d}, Measured intensity autocorrelation of the femtosecond mode-locked state, consistent with a $\sim$750\,fs pulse duration. The reconstructed autocorrelation of the filtered simulated spectrum (blue line) and filtered experimental spectrum (pink dashed line).
        \textbf{e}, Measured and simulated spectra of the femtosecond mode-locking regime. Shading marks the content above 1570 nm that lies beyond the pulse shaper's bandwidth and is excluded from the autocorrelation in (d). Scale bar, 150 \micro m \textbf{(a)}.
    }    
    }
    \label{Fig4:Femto}
\end{figure*}
\noindent  
We first characterize the QNLI-based laser in the self-starting regime with the setup shown in Fig.\,\ref{Fig3:PMLL}a.
We bias the QNLI using the on-chip heater to maximize its modulation depth (SI section I, Figs.~S1,~S2) and electrically drive the RSOA with a direct current source. 
We collect the laser's output with a lensed-tip fiber and analyze it spectrally with an optical spectrum analyzer (OSA) and temporally via autocorrelation and radio-frequency (RF) measurements. 
We progressively increase the RSOA drive current until a sech-squared spectrum spontaneously appears at the OSA. To test the robustness of this regime, we toggle the RSOA on and off across multiple power cycles. Following a short transient, the laser repeatedly converges to the same DQS state (Fig.\,\ref{Fig3:PMLL}b), exhibiting robust, self-starting behavior, and confirming turn-key operation.
The experimental trace in the steady state and the lumped-element simulation (SI section VI.B, Table 2) are plotted in Fig.\,\ref{Fig3:PMLL}c. 
The agreement between the two indicates the formation of a single 4.9\,ps-long DQS (inset).

Guided by the predictions of Fig.\,2f, we repeat the same measurement with a different phase bias. As anticipated, this drives the laser into harmonic mode-locking. The corresponding optical spectrum and autocorrelation are shown in Fig. \ref{Fig3:PMLL}e and Fig.\,\ref{Fig3:PMLL}f, respectively. These results confirm the formation of four DQSs with dispersion-compensated durations of $\tau_s$ = 4.27\,ps and separated by $\sim$22\,ps. 
We measure fiber-coupled power of 393 \micro W, yielding a pulse energy of $\sim$8.7 fJ in the fiber corresponding to $\sim$35 fJ on chip, confirming ultra-energy-efficient mode locking of the laser.

Finally, to confirm the quadratic frequency comb coherence, we proceed to beat note measurements. The free-running self and heterodyne beatnotes have linewidths of 16.7\,kHz and 2.64\,MHz, respectively, as shown in Fig.\,\ref{Fig3:PMLL}d and the inset of Fig.\,\ref{Fig3:PMLL}e (SI section VIII, Fig.~S9). These linewidths are sufficiently narrow to fall within the bandwidth requirements for active electronic feedback, and thus are consistent with stabilization schemes for achieving a fully stabilized comb source\,\cite{ye_femtosecond_2005}.

\section*{Femtosecond operation} 
To further shorten the generated pulse length of the MLL, in a second device, we replace the heater with a high-speed electro-optic modulator (EOM), and drive it with an RF input at twice the cavity free spectral range, as shown in Fig.\,\ref{Fig4:Femto}a (Methods). We first study this operation regime numerically (Fig.\,\ref{Fig4:Femto}b,c), for which Eqs.\,\eqref{eq:qnli} remain valid, provided that $\varphi \mapsto \varphi(\tau) = M\,\cos(\omega_m\tau)$ where $M$ is the EOM modulation depth and $\omega_m$ is the modulation frequency. 
In the presence of the RF modulation, the system converges toward a Gaussian pulse, as expected for an amplitude-modulated active MLL (Fig.\,\ref{Fig4:Femto}b, gray). However, the role of the nonlinear phase shift in the $\chi^{(2)}$ section of the QNLI becomes substantial and significantly reduces the overall mode-locked pulse duration (Fig.\,\ref{Fig4:Femto}b, light blue). The pulses follow the theoretical soliton condition similar to Fig.\,\ref{Fig2:Simulations}b (SI section IX, Fig.~S10), confirming operation in the DQS regime. The nonlinear response of the QNLI, combined with the strong electro-optic effect of TFLN, produces a flat-top spectrum\,(Fig.\,\ref{Fig4:Femto}c) with chirped intracavity pulses in the picosecond regime ($\sim$1.5 ps), which, upon dispersion compensation, results in femtosecond pulse durations (Fig.\,\ref{Fig4:Femto}b, dark blue).

To characterize the device, we apply a $\sim$14.2 GHz RF tone to the intracavity EOM. Figure\,\ref{Fig4:Femto}d and\,\ref{Fig4:Femto}e show the measurement results for the femtosecond operation of the QNLI-based MLL. The laser produces 400 \micro W of in-fiber power, yielding a 28.2 fJ pulse energy, corresponding to $\sim$113 fJ on chip. In this regime, the measured spectrum (Fig.\,\ref{Fig4:Femto}e) spans over 8 nm at the 10 dB level, largely following the simulated profile. The difference between the measured and simulated spectra on the long-wavelength side of the spectrum may arise from the interplay of birefringence-induced polarization mixing and Michelson interference, producing wavelength-dependent cavity feedback. This can be mitigated through optimizing the waveguide geometry, which could increase the usable comb bandwidth and yield significantly shorter pulse durations.

While the QNLI laser natively generates a broad spectrum extending beyond 1570 nm, temporal characterization was constrained by the operating bandwidth of the programmable pulse shaper used for external dispersion compensation. Despite this spectral truncation, the autocorrelation trace reveals a $\sim$750 fs pulse (Methods). To verify this, we applied the experimental spectral filter ($<$1570 nm) to both the measured and simulated optical fields. As shown in Fig.\,\ref{Fig4:Femto}d, the resulting pulse reconstructions closely resemble the measured autocorrelation trace. The agreement between experiment and simulation supports a $\sim$750\,fs pulse duration arising from the compensated band and is consistent with intracavity DQS formation.

\section*{Discussion and outlook}
Our results, to the best of our knowledge, mark the first demonstration of an integrated mode-locked laser in the quadratic soliton regime. We also introduce a QNLI, which in combination with a semiconductor gain element yields an electrically driven femtosecond mode-locked laser on thin-film lithium niobate nanophotonics.
The QNLI offers a near-instantaneous effective saturable absorber beyond what could be achieved in other nonlinear photonic platforms in terms of energy-efficiency and speed (SI section V, Fig.~S6).
We experimentally demonstrate turn-key soliton-mode locking with semiconductor gain, deep in the normal dispersion regime, and do not observe undesired nonlinear effects in the semiconductor gain element. A modified CGLE confirms dissipative quadratic soliton formation and is consistent with measurements.
The laser performance can be further enhanced towards self-starting $\sim$250-fs operation (SI section VII.C, Fig.~S7b) by utilizing broadband couplers\,\cite{moille_broadband_2019}, multimode interferometers\,\cite{chen_design_2021} or apodized gratings\,\cite{kartner_design_1997,yu_photonic-crystal-reflector_2019}, as well as mitigating unwanted mode-mixing which currently limits the operation bandwidth. 
While our current results focus on ultra-energy-efficient laser mode locking, the pulse energy and output power of the laser can be substantially increased. We numerically confirm that peak powers greater than 1 W can be achieved on chip (SI section X, Fig.~S11, Table 3), compatible with octave-spanning spectral broadening on the same platform \cite{sekine_multi-octave_2025} and sufficient for f-2f interferometry, providing a path towards self-referenced chip-scale comb sources. 

The presented electrically pumped QNLI-based MLL can drive the wealth of ultrafast optical functionalities in LN monolithic ultrafast PICs without the need for external laser sources.
Our ultra-energy-efficient quadratic soliton mode-locking scheme can be applied to other wavelength ranges within the entire TFLN transparency window, making it compatible with other gain media, either semiconductor or solid-state\,\cite{liu_photonic_2022,yang_titaniumsapphire--insulator_2024,zeng_integrated_2026}, whether they operate in the normal dispersion regime or not. 

\renewcommand{\bibsection}{\section*{References}} 
\bibliographystyle{ieeetr}
\bibliography{lib}

\section*{Methods}

\noindent\textbf{Device fabrication}
\\
The QNLI was fabricated on 700-nm-thick X-cut MgO-doped thin-film lithium niobate on a SiO$_2$/Si substrate (NANOLN). Quasi-phase matching is achieved for the QNLI device(s) used as a passive (hybrid) mode-locked laser with a 1.2-mm-long (4.6-mm-long) region through periodic poling via Cr electrodes. After poling, electron beam lithography is used to pattern the nominally 1\,\micron{} wide waveguides using hydrogen silsesquioxane (HSQ) as the resist. Next, the TFLN is etched 350\,nm deep using Ar\textsuperscript{+} plasma, leaving a $\sim$$60^\circ$ sidewall angle.  Plasma-enhanced chemical vapor deposition is used to deposit 800\,nm of SiO$_2$ cladding. The heaters and RF electrodes are then patterned on top of the cladding using electron beam lithography, this time using PMMA as the resist. Electron-beam evaporation is then used to deposit 10\,nm of Cr, followed by 250\,nm of Au. The RF coplanar waveguide, designed to achieve a 50\,$\Omega$ impedance, implements 28\,\micron{} / 8\,\micron{} / 28\,\micron{} (GSG) electrode widths, an electrode gap of 4\,\micron{}, and a length of 3.5\,mm. The heaters implement a conductor of 4\,\micron{} width and 2\,mm path length, with a measured resistance of 100\,$\Omega$. Finally, the waveguide facets are mechanically polished to improve facet coupling. The footprint of the QNLIs used for passive and hybrid mode-locking are 0.7\,mm~×~3.5\,mm and 0.7\,mm~×~7\,mm, respectively.\\

\noindent\textbf{Experimental setup for passive mode-locking}
\\
The QNLI is affixed to a stationary post, where an RSOA is aligned to the device facet with a 6-axis nanopositioning stage, forming a Fabry-Perot cavity between the highly reflective ($>$90\%) normal facet of the C-band single-angled facet RSOA and the QNLI. We use a Thorlabs RSOA (SAF1126C) for all results except Fig.\,\ref{Fig3:PMLL}d, in which we use a Seminex RSOA (COC-286-20). A mode-matching taper enables efficient coupling between the device and the RSOA. A laser diode controller (Thorlabs LDC240C) supplies current to the RSOA. A function generator (Agilent 33220A) supplies power to the on-chip heaters to control the fixed phase bias in the QNLI. The laser output is coupled into a lensed fiber ($\sim$6 dB chip-to-fiber coupling loss) that is then split between an optical spectrum analyzer (Yokogawa AQ6370D), and a programmable wave shaper (Coherent WS-01000A-C-S-1-AA-01). The output of the wave shaper is then amplified by an erbium-doped fiber amplifier (Thorlabs EDFA100P), after which a 90:10 coupler sends a majority of the amplified signal to an intensity autocorrelator (Femtochrome FR-103XL) and the rest to a heterodyne detection scheme. For the autocorrelation measurement in Fig.\,\ref{Fig3:PMLL}f, we apply +7 ps/nm of dispersion compensation. Beat note measurements are performed by combining a local oscillator from a CW laser (Toptica CTL 1550) with the comb output, which was then passed to a fast photodiode (Thorlabs DX30AF) whose electrical signal was detected by the electrical spectrum analyzer (Rohde \& Schwarz FSW). \\

\noindent\textbf{Experimental setup for femtosecond mode-locking}
\\
The femtosecond results use a second device with a 4.6-mm-long poled section, in which the heater is replaced by the 3.5-mm-long coplanar waveguide phase modulator along the unpoled arm (Fig.\,\ref{Fig4:Femto}a). The output measurement scheme is identical to that of the passive mode-locked laser experiment. However, here, an RF signal from a signal generator (Rohde \& Schwarz SMA100B) is amplified by a high-power RF amplifier (Mini-Circuits ZVE-3W-183+) before being applied to the coplanar waveguide phase modulator on the device. An RF power meter (Ladybug LB680A) performed an RF power measurement and provided a 50 $\Omega$ termination to the transmission line. The RF frequency was tuned to match the second harmonic of the cavity FSR, resulting in harmonic mode-locking. For the autocorrelation measurement in Fig.\,\ref{Fig4:Femto}d, we apply -0.6 ps/nm of dispersion compensation.

The spectra in Fig.\,\ref{Fig4:Femto}e were measured before the wave shaper. Since the wave shaper transmission drops off after 1570 nm, the spectral content beyond this wavelength does not reach the autocorrelator. Thus, we reproduce the autocorrelation from both the measured and the simulated spectra. To do this, we first remove the spectral content above 1570\,nm. We then apply a quadratic spectral phase to each truncated spectrum and recover the pulse intensity by Fourier transforming the spectral field. Finally, we compute the autocorrelation of each pulse and compare it with the measured autocorrelation in Fig.\,\ref{Fig4:Femto}d. The dispersion was fitted to the measured autocorrelation, giving $|D| = 0.116$\,ps/nm and $0.121$\,ps/nm for the measured and simulated spectra, respectively. 
This small residual dispersion is required to reproduce the measured autocorrelation width, indicating that the $-0.6$\,ps/nm applied by the pulse shaper did not fully compensate the dispersion of the pulse. The measured autocorrelation deviates slightly from both reconstructions in the wings, which we attribute to higher-order spectral phase not captured by a single dispersion term. The two spectral reconstructions give pulse durations of 768\,fs and 776\,fs for the measured and simulated spectra, respectively. The measured autocorrelation width of 1.14\,ps corresponds to a pulse duration of 739\,fs assuming a sech$^2$ deconvolution factor of 1.543. Considering these estimates, we report a pulse duration of approximately 750\,fs.\\

\noindent\textbf{Data availability} 
\\
The datasets used in the current study are available from the corresponding author upon request.\\

\noindent\textbf{Code availability} 
\\
The code used in the current study is available from the corresponding author upon request. 

\vspace{-10pt}
\section*{Acknowledgements} 
\vspace{-10pt}
\noindent The device nanofabrication was performed at the Kavli Nanoscience Institute (KNI) at Caltech. We thank Lin Yi for lending equipment. We thank Auro M. Perego for fruitful discussions. We acknowledge support from DARPA award D23AP00158, NSF grant nos. 2408297 and 2625394, AFOSR award FA9550-23-1-0755, ARO grant W911NF-26-1-A311, the Center for Sensing to Intelligence at Caltech, the Alfred P. Sloan Foundation, and NASA/JPL. N.E. acknowledges support from the Belgian American Educational Foundation (B.A.E.F.), the SofinaBo\"{e}l Fund and the Horizon Europe research and innovation programme of the European Union under the Marie Sk\l{}odowska--Curie grant agreement no. 101103780.

\vspace{-10pt}
\section*{Author contributions} 
\vspace{-10pt}
\noindent B.K.G., N.E. and A.M. conceived the project. B.K.G. designed the device with input from N.E\@. S.Z. fabricated the device with help from M.S\@. B.K.G., N.E. and Y.X. performed the experiments, with assistance from M.B\@. N.E. and Y.X. carried out the numerical simulations. B.K.G., N.E. and A.M. wrote the manuscript with input from all authors. A.M. supervised the project.

\vspace{-10pt}
\section*{Competing interests}
\vspace{-10pt}
\noindent B.K.G., N.E., and A.M. are inventors on a patent application (US application no. US19/072,822 and International application no. PCT/US2025/018794). A.M. is an inventor on the US patent no. 12368278. A.M. is involved in developing photonic integrated nonlinear circuits at PINC Technologies Inc. A.M. has an equity interest in PINC Technologies Inc. The remaining authors declare no competing interests.

\end{document}